# Dispersal-induced resilience to stochastic environmental fluctuations in populations with Allee effect

**Rodrigo Crespo-Miguel**[1,*], **Javier Jarillo**[2,**], **Francisco J. Cao-García**[1,3,***]

[1] Departamento de Estructura de la Materia, Física Térmica y Electrónica, Facultad de Ciencias Físicas, Universidad Complutense de Madrid. Plaza de Ciencias 1, 28040 Madrid, Spain

[2] Research Unit of Environmental and Evolutionary Biology, Namur Institute of Complex Systems, and Institute of Life, Earth, and the Environment, University of Namur, Rue de Bruxelles 61, Namur, 5000, Belgium.

[3] Instituto Madrileño de Estudios Avanzados en Nanociencia (IMDEA-Nanociencia). Calle Faraday 9, 28049 Madrid, Spain.

* Email: rodcresp@ucm.es

** Email: javier.jarillodiaz@unamur.be

*** Email: francao@ucm.es. **Corresponding author**.

**Abstract**

Many species are unsustainable at small population densities (Allee Effect), i.e., below a threshold named Allee threshold, the population decreases instead of growing. In a closed local population, environmental fluctuations always lead to extinction. Here, we show how, in spatially extended habitats, dispersal can lead to a sustainable population in a region, provided the amplitude of environmental fluctuations is below an extinction threshold. We have identified two types of sustainable populations: high-density and low-density populations (through a mean-field approximation, valid in the limit of large dispersal length). Our results show that patches where population is high, low or extinct, coexist when the population is close to global extinction (even for homogeneous habitats). The extinction threshold is maximum for characteristic dispersal distances much larger than the spatial scale of synchrony of environmental fluctuations. The extinction threshold increases proportionally to the square root of the dispersal rate and decreases with the Allee threshold. The low-density population solution can allow understanding difficulties in recovery after harvesting. This theoretical framework provides a novel approach to address other factors, such as habitat fragmentation or harvesting, impacting population resilience to environmental fluctuations.

## I. Introduction

Many species need a minimum population density to be viable such as the island fox [1], the polar bear [2], the american ginseng [3], and the atlantic codfish [4] among others. This minimum viable population density is named Allee threshold [5], and below it, the population declines towards extinction, a phenomenon called the (strong) Allee effect.

Field researches have characterized the impact for plants and animals of reaching population densities below the Allee threshold. The strength of the impact depends on the strength of the Allee Effect [6], the presence of harvest [7,8], and the absence of positive human intervention [9,10]. Many of these depleted populations never recover and become extinct in some years. Some depleted populations take many years (much more than the average lifetime of the species) to get out of this situation and eventually recover. In animals, monogamous species with long lifetimes are more likely to show Allee Effect [1,11], in addition to solitary species with difficulties for finding a breeding mate or an unbalanced male-female ratio [2]. Other causes that explain the Allee effect in animals are non-efficient feeding [12], difficulties to survive in an environment with predators, competitors, or human harvest [13,14], or inbreeding depression [15,16]. In plants, less efficient pollination or fruit production (decreasing at small populations) [3] seem to be the principal causes providing Allee Effect. Most of the articles cited above qualitatively describe how a particular species in a low population density situation has difficulties surviving due to Allee Effect.

Theoretical papers have addressed the general question of the eradication of alien species [17], and of the spatial patterns influence on the spread of invading species [18], observed in gypsy moth [19]. These papers show that both the Allee Effect and environmental variability can contribute to the extinction of a population. Other works have studied the effects of stochasticity in species vulnerable to the Allee Effect [20–23] describing mean time to extinction, or probability of extinction after a given time, for a single location. Migration between locations increases the mean time to extinction, as it has been shown in a metapopulation model with a 3x3 grid [24]. Studies of a locally endangered butterfly sustain the critical role of immigration in the regional dynamics to counterbalance the Allee effects [25]. Recently, Dennis et al. [26] have shown that an external constant migration term can sustain the population in the presence of stochastic environmental fluctuations. Here, we go a step further and show that a spatially extended population with dispersal between the locations can be sustainable. We compute the stationary population probability distributions in the mean-field limit (large dispersal distance), which elucidates the spatially extended population dynamics for finite dispersal distance. These results clarify the conditions for sustainability in spatially extended habitats, quantifying the effects of dispersal in the resilience to stochastic environmental fluctuations.

We have studied a one-dimensional model, which is a good approach for some ecological systems such as rivers or oceanic water columns [27]. One-dimensional models also allow a more straightforward yet accurate study of many characteristics of interaction-based population dynamics [28].

The results we present here provide insight into how natural or human-induced changes in the species' dynamical parameters would influence its extinction risk due to environmental fluctuations. In particular, they provide information on how an increase in the amplitude of environmental fluctuations can affect the sustainability of a population. This problem is of particular present relevance as several regions of the Earth are increasing its climate variability [29].

## II. Spatially extended population model

We introduce a spatially extended, one-dimensional population model, including Allee effects, environmental fluctuations, and dispersal. This model allows us to assess the resilience of populations to environmental fluctuations and the role played by dispersal in this resilience.

The deterministic Allee model [30,31] gives the local deterministic dynamics of the population density $N(x,t)$ at location x and time t. This dynamics is determined by a characteristic extinction rate $r$, a carrying capacity $K$ (stable, viable population density), and an Allee threshold $A$ (minimum viable population density). Note that the population density $N(x,t)$ is defined as the local number of individuals per unit of length at a given time. Additionally, environmental stochasticity is introduced through an additional stochastic contribution, $\sigma N\, dB$, proportional to the population density [32]. The amplitude of these environmental fluctuations is given by $\sigma$, and $dB(x,t)$ is a normalized Gaussian random field with a spatial scale of synchrony $l_e$, giving the spatial scale of synchrony of the environmental fluctuations, which is the characteristic distance at which environmental fluctuations remain correlated [33]. Therefore, the local dynamics of a population density N(x,t) in the stochastic Allee model is given by

$$dN(x,t)|_{local} = r\,N(x,t)\left(\frac{N(x,t)}{A} - 1\right)\left(1 - \frac{N(x,t)}{K}\right) dt + \sigma\,N(x,t)\,dB(x,t). \quad (1)$$

The first term corresponds to the deterministic Allee model [18,30,31]. This equation implies a rate of return to extinction for populations close to extinction of $\gamma_0 = r$ and a rate of return to the carrying capacity for populations close to the carrying capacity of $\gamma_K = r\,(K/A-1)$. The second term in Eq. (1) gives the contribution of stochastic environmental fluctuations to the changes in the local population, with an amplitude of the environmental fluctuations $\sigma$. The random field $dB(x,t)$ is given by increments of standard Brownian motions in each position with zero mean and variance $dt$, and it is spatially correlated with an exponential autocorrelation of length $l_e$, which is the spatial scale of synchrony of environmental fluctuations [33].

The dispersal couples the dynamics in the different locations. We consider that individuals disperse away with a rate $m$ to a characteristic distance $l_m$. Thus, dispersal gives an additional contribution to the dynamics of

$$dN(x,t)|_{dispersal} = -m\,N(x,t)\,dt + m\,dt \int N(x-y,t)\,f(y)\,dy \quad (2)$$

which makes the dynamics non-local. The first term represents the population decrease at position $x$ due to individuals that disperse away with probability $m\,dt$. The second term gives the population increase due to individuals that disperse to position $x$ from a position displaced a distance $y$ where the population is $N(x-y,t)$. Therefore, $m$ is the rate of random dispersal to a position at a distance $y$ with probability $f(y)$, where $f(y)$ has been taken as a Gaussian with variance $l_m^2$ and zero mean. Dispersal rate $m$ is the same in both terms in Eq. 2 because our model assumes no external migration: Every individual leaving a patch moves to another within the ecosystem, whereas every individual arriving at a patch must have come from the same ecosystem. Furthermore, we consider a homogeneous habitat, so neither the dispersal rate $m$ nor the dispersal profile $f(y)$ (nor any other parameter of the model) depends on position $x$. Hence, individuals disperse at a rate $m$ to typical distances of the order of $l_m$. As the dispersal term in Eq. (2) is proportional to the population density $N(x,t)$, depleted regions will receive a net population flux from nearby non-depleted regions.

The combination of local and dispersal contributions to the change in population density gives the complete spatially extended dynamics,

$$dN = dN|_{local} + dN|_{dispersal} \qquad (3)$$

Typical late time population distributions given by this dynamical equation can be seen in Fig. 1 (panels A, C, and E). Table 1 gives a summary of the variables and parameters used and their units.

| Variables | Description |
|---|---|
| $N(x,t)$ | Population density at a given position x and a time t. Units of length$^{-1}$. |
| $A$ | Allee threshold of the species; below it, the species has negative growth. Units of length$^{-1}$. |
| $K$ | Carrying capacity of the species, meaning stable, viable population density. Units of length$^{-1}$. |
| $r$ | Extinction rate (at very low population densities). Units of time$^{-1}$. |
| $\gamma_0$ | Rate of return to extinction. Units of time$^{-1}$. |
| $\gamma_K$ | Rate of return to carrying capacity. Units of time$^{-1}$. |
| $m$ | Dispersal rate of the species. Units of time$^{-1}$. |
| $l_e$ | Spatial scale of synchrony of environmental fluctuations. Units of length. |
| $l_m$ | Mean distance traveled by the dispersed individuals (characteristic width of the Gaussian dispersion function). Units of length. |
| $\sigma$ | Amplitude of the environmental fluctuations, gives the standard deviation of the environmental fluctuations. Units of time$^{-1/2}$. |
| $\sigma_{extinction}$ | Extinction threshold for the amplitude of environmental fluctuations (minimum amplitude of the fluctuations that ensures global extinction). Units of time$^{-1/2}$. |

*Table 1: Variables and parameters used in this article (definitions and units).*

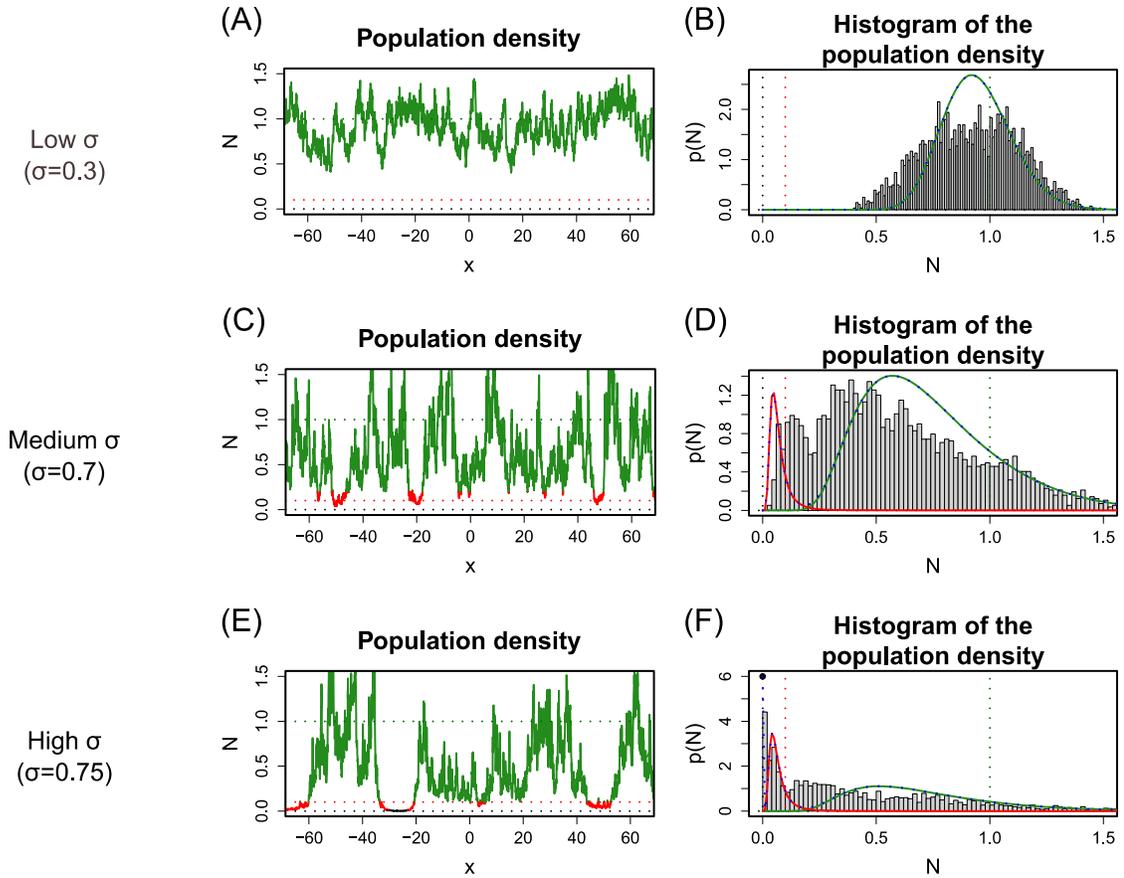

*Figure 1: Spatial profiles of population density, their associated averaged population density histograms compared with mean-field population probability distributions.* Panels A, C, and E: spatial profile of the population density for simulation at a long time, t=1000, with extinction rate r=0.1, Allee threshold A=0.1, carrying capacity K=1, dispersal rate m=1, and dispersal distance equal to the spatial scale of synchrony of environmental fluctuations, $l_m=l_e=1$. Panels B, D, and F: population density histograms for the spatial profiles of population densities shown in Panels A, C, and E, respectively. The curves in Panels A, C, and E show the patches in extinction (black) high-population (green), and low-population states (red), according to the histograms in Panels B, D, and F. Panels B, D, and F also show the fit to a linear combination of the mean-field population probability distributions. The result for this fits are $p(N)=p_{high}(N)$ in Panel B, $p(N)=0.08p_{low}(N)+0.92p_{high}(N)$ in Panel D and $p(N)=0.06p_0(N)+0.22p_{low}(N)+0.72p_{high}(N)$ in Panel F. Each contribution is represented with its fitted weight. $p_{low}(N)$ (red line) and $p_{high}(N)$ (green line) correspond, respectively, to the low and high-density mean-field population probability distribution solutions. They are given by the two

*nonzero branches of solutions of the mean-field equations for values of σ below the extinction threshold (see also Fig. 2). $p_0(N)$ (black point) is the zero population density solution (i.e., extinction). Red dashed lines indicate Allee threshold value A=0.1, green dashed lines indicate carrying capacity K=1 and black dashed lines indicate N=0. Note the similarities between the population density histograms obtained from direct numerical simulation and the fit to the linear combination of the population probability distributions obtained with the mean-field limit approximation (i.e., the large dispersal distance limit, $l_m/l_e \to \infty$). However, the real histograms are displaced to the left due to the border effects, which are effects beyond the mean-field approximation (i.e., due to finite dispersal distance, $l_m/l_e=1$). Note also that increasing the environmental fluctuations σ increases the presence of regions where the population is depleted or extinct, as σ approaches the extinction threshold ($\sigma_{extinction}=0.80$ for the parameter values in this figure).*

### III. Numerical simulations

The numerical simulations of the previously described dynamical equation is performed taking the scale of synchrony of environmental fluctuations, $l_e$, as reference length, i.e., $l_e=1$, and 20 lattice nodes per unit length. The total length of the simulation box was 140 times the maximum of $l_e$ and $l_m$, and we consider periodic boundary conditions (aiming to obtain results for infinite habitat). The time resolution was 50 times smaller than the minimum of the characteristic times of the dynamics (i.e., the minimum of the inverses of the rates $r$ and $m$). These resolutions, simulation boxes, and boundary conditions guarantee that the dynamics is well-resolved (in time and space) and mimics an infinite habitat for better comparison with the results found with the mean-field approximation [34,35] described below. In this way, we performed numerical simulations of spatially extended populations, starting from a population density equal to the carrying capacity in each node of the simulation box. We ran several simulations for each set of parameters with different amplitudes of the environmental fluctuations. See Appendix C for further details on the simulation algorithm.

We define the extinction threshold, $\sigma_{extinction}$, as the characteristic amplitude of environmental fluctuations above which the environmental fluctuations lead to the global extinction of population. The environmental fluctuation extinction threshold $\sigma_{extinction}$ is obtained from long time simulation ($t = 1000 = 100 \, r^{-1}$) as the center of the transition interval from the never extinct to the always global extinct final state.

Besides, relevant information of a spatially extended population is how probable it is to find a given population density in a given location, *i.e.*, to determine the population probability distribution. We have assumed homogeneous habitat conditions, represented by location-independent population dynamics parameters (extinction rate *r*, carrying capacity *K*, Allee threshold *A*, and amplitude of environmental fluctuations $\sigma$). Thus, the population probability distribution *p(N)* does not depend on location and gives the probability to find the population density *N* in any site. The population probability distribution *p(N)* is computed from numerical simulations doing population density histograms, like those shown in Fig. 1 Panels B, D, and F.

### IV. *Analytical population probability distribution*

Additionally, we can get further insight into the population dynamics through a more analytical approach to the computation of the population probability. The stochastic differential equations for the stochastic Allee model with and without dispersal, Eqs. (1) and (3), have the form $dN = F(N) \, dt + \sqrt{v(N)} \, dB$. For equations of this form, if a stationary population probability distribution exists, it is given by [36]

$$p(N) = \frac{n}{v(N)} \cdot \exp\left(2 \int \frac{F(N)}{v(N)} dN\right), \qquad (4)$$

where *n* is a normalization factor.

Therefore, for the stochastic Allee model without dispersal, Eq. 1, the stationary population probability distribution is

$$p(N) = \frac{n \cdot \exp\left(\frac{1}{\sigma^2}\left(\frac{2rN}{K} + \frac{2rN}{A} - \frac{rN^2}{AK}\right)\right)}{\sigma^2 N^{2+2r/\sigma^2}}. \qquad (5)$$

This population distribution has a divergence in $N = 0$ and is not normalizable, which means that a population with an Allee-type growth will always become extinct in the absence of dispersal ($m = 0$). (The extinction is faster for larger environmental fluctuations. See Appendix B.)

Only populations with dispersal may be stable in the long term. For the stochastic Allee model with dispersal, Eq. (3), the stationary population probability distribution with dispersal is

$$p(N) = \frac{n \cdot \exp\left(\frac{1}{\sigma^2}\left(\frac{2rN}{K} + \frac{2rN}{A} - \frac{rN^2}{AK} - \frac{2mI}{N}\right)\right)}{\sigma^2 N^{2 + 2(r+m)/\sigma^2}}, \quad (6)$$

where the coupling term $I(x) = \int N(x-y) f(y) \, dy$ makes the population probability in one location depending on the values of the population density in the surrounding region. This additional dispersal term makes the population distribution normalizable and with no-zero mean for certain ranges of values of $m$ and $\sigma$.

## V. *Mean-field approximation*

We propose here to combine the analytic expression in Eq. (6) with the mean-field approximation to deal with the coupling term $I(x)$. When the dispersal length is much larger than the spatial scale of synchrony of environmental fluctuations, $l_m \gg l_e$, (for example, the case of long-distance migrant birds), the mean-field approximation is a good approximation. The mean-field approximation assumes the stationary population probability is approximately equal in all points of space. (Long-distance dispersal, $l_m \gg l_e$, makes the population density more homogeneous, recolonizing low-populated points from those that are "overpopulated".) The mean-field approximation implies that the coupling term $I$ in Eq. (6) can be treated as position-independent, $I(x) = I$, and we can approximate it by the mean value of the population density

$$I = \int_0^\infty N \, p(N) \, dN. \quad (7)$$

This approximation mimics the dynamics of long-distance dispersal (see Fig. A2 in Appendix A). The extinction threshold, $\sigma_{extinction}$, was defined here as the amplitude of environmental fluctuations above which the environmental fluctuations leads to the global extinction of the population. In the mean-field approximation, the extinction threshold can be computed directly obtaining the value of the environmental amplitude where there is no longer a solution of the system of equations formed by Eqs. (6) and (7) (*i.e.*, the point where the green and the red curve merge in Fig. 2A), except the extinction in all space solution (non-normalizable divergence in N=0). Thus, when the amplitude of environmental fluctuations exceeds the extinction threshold, the only possible solution implies global extinction (we find population at N=0 for long times with probability one, $p(N = 0) = 1$), and it is not normalizable. Therefore, the mean-field approximation allows us to compute the stationary probability distribution and to estimate the extinction threshold for a particular set of parameters (which is a close upper limit of the real extinction threshold when the dispersal distance of the species is large enough). We get two branches of solutions, which represent two different equilibria: the high mean population density solution, $p_{high}$, and low mean population density solution, $p_{low}$. See Figs. 1 and 2.

The diagram of the mean population densities $I$ for the two branches of solutions depends on the parameters of the model (Fig. 2). Lower Allee thresholds $A$ displace the diagram towards higher amplitudes of environmental fluctuations $\sigma$, but lower the mean population $I$ and make more prominent the minimum of the lower branch. Higher dispersal rates $m$ displace the diagram to higher amplitudes of environmental fluctuations $\sigma$ and slightly to higher mean population densities $I$, remarking the role of dispersal as a stabilizing factor of the population. (See Fig. 2 and Fig. A1 in Appendix A.)

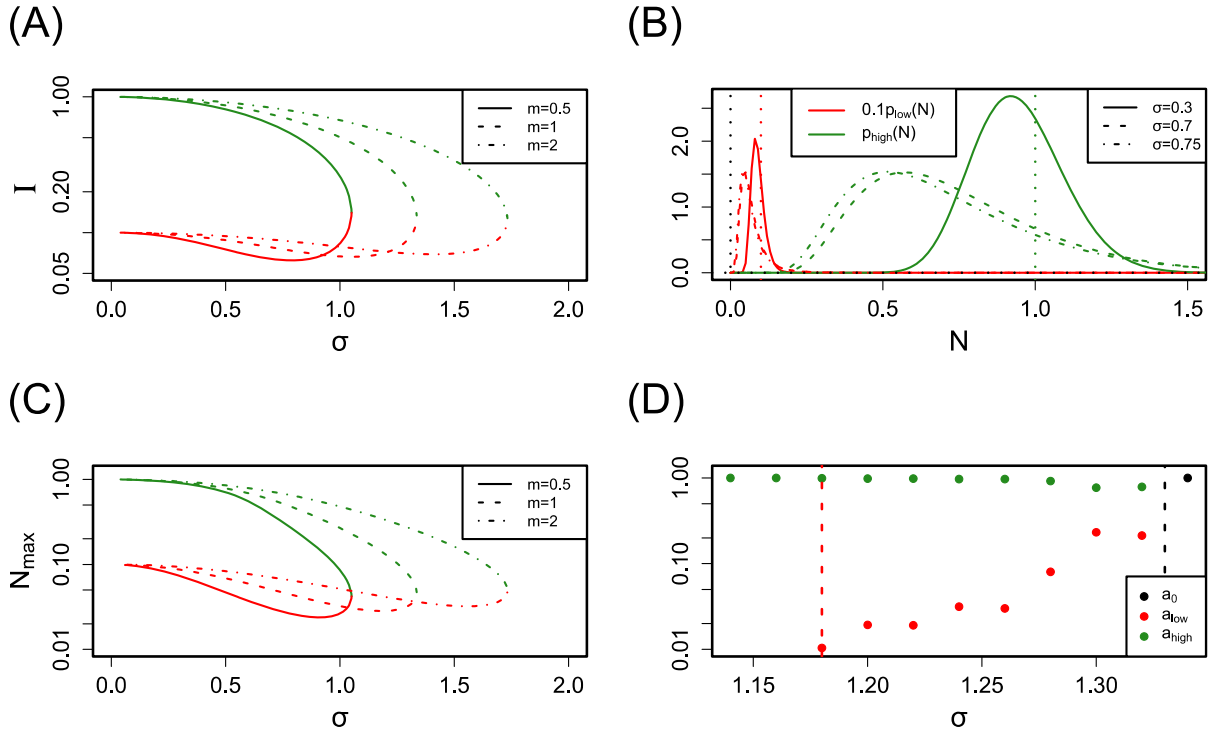

*Figure 2: Solutions in the mean-field approximation: (Panel A) Mean population density I as a function of environmental noise amplitude σ for the two nonzero branches of solutions at the*

*mean-field approximation ($l_m \gg l_e$): high-density (green) and low-density (red) branches (shown in logarithmic scale). Extinction rate r=0.1, carrying capacity K=1, Allee threshold A=0.1, and dispersal rate m=0.5 (solid line), m=1 (dashed line) and m=2 (dash-dotted line). (Panel B) High-density (green) and low-density (red, divided by 10 to make the figure more visible) population probability distributions, $p_{high}$ and $p_{low}$, respectively. They are calculated at the mean-field approximation for the same parameters of Panel A and m=1, for values of the environmental fluctuations amplitude σ=0.3 (solid line), σ=0.7 (dashed line) and σ=0.75 (dash-dotted line). Vertical lines show N=0 (black), N=A=0.1 (red) and N=K=1(green). (Panel C) Position of the maximum of mean-field distributions $N_{max}$ as a function of environmental noise amplitude σ, for the two nonzero branches of solutions at the mean-field approximation (shown in logarithmic scale) with the same parameters as in Panel A. (Panel D) Fitted contribution of each mean-field distribution in stationary distributions simulated by population dynamics assuming mean-field limit (see appendix C) such as $p(N)_{simulated} = a_{low}\, p_{low}(N) + a_{high}\, p_{high}(N) + a_0 p_0(N)$, where $p_0(N)$ is the zero population density solution (i.e., extinction) and $a_{low} + a_{high} + a_0 = 1$, for extinction rate r=0.1, carrying capacity K=1, Allee threshold A=0.1, and dispersal rate m=1. Vertical red dashed line show the amplitude of environmental fluctuations which gives the position of the minimum in the lower branch of $N_{max}$ (panel C), and vertical black dashed line shown the extinction threshold $σ_{extinction}$ calculated in the mean field limit. The vertical axis is shown in logarithmic scale and values below 0.01 are not represented. The figure shows that contributions of $p_{low}(N)$ start to appear for sigmas equal or greater than the position of the minimum in the lower branch of $N_{max}$. Additionally extinction appears for amplitudes of environmental fluctuations greater than $σ_{extinction}$.*

## VI. Maximum approximation

The maximum approximation assumes the extinction threshold is the value of the amplitude of environmental fluctuations σ that locates the maximum of $p_I(N)$ [where $p_I(N)$ is is the population probability for a given *I*] at the Allee threshold A. Adding the maximum approximation allows a quick estimation of the extinction threshold,

$$\sigma_{extinction} = \sqrt{m\left(\frac{I}{A} - 1\right)}. \qquad (8)$$

This expression points that the extinction threshold approximately increases with the square root of the dispersal rate and decreases with the Allee threshold A (Fig. 3). To obtain the value of $\sigma_{extinction}$ estimated with the mean-field and maximum approximations, we must simultaneously solve numerically Eqs. (6), (7), and (8). It should be noted that the maximum approximation is only an additional approximation, which can be added to the method introduced in the previous section [$\sigma_{extinction}$ is the value of the environmental amplitude where there is no longer a solution for the set of equations Eqs. (6) and (7)]. The maximum approximation allows a faster estimation of the extinction threshold and shows that the extinction threshold approximately grows with the root of the dispersal rate m. However, the method described in the previous section is more accurate and preferred to compute the extinction threshold in the mean-field limit.

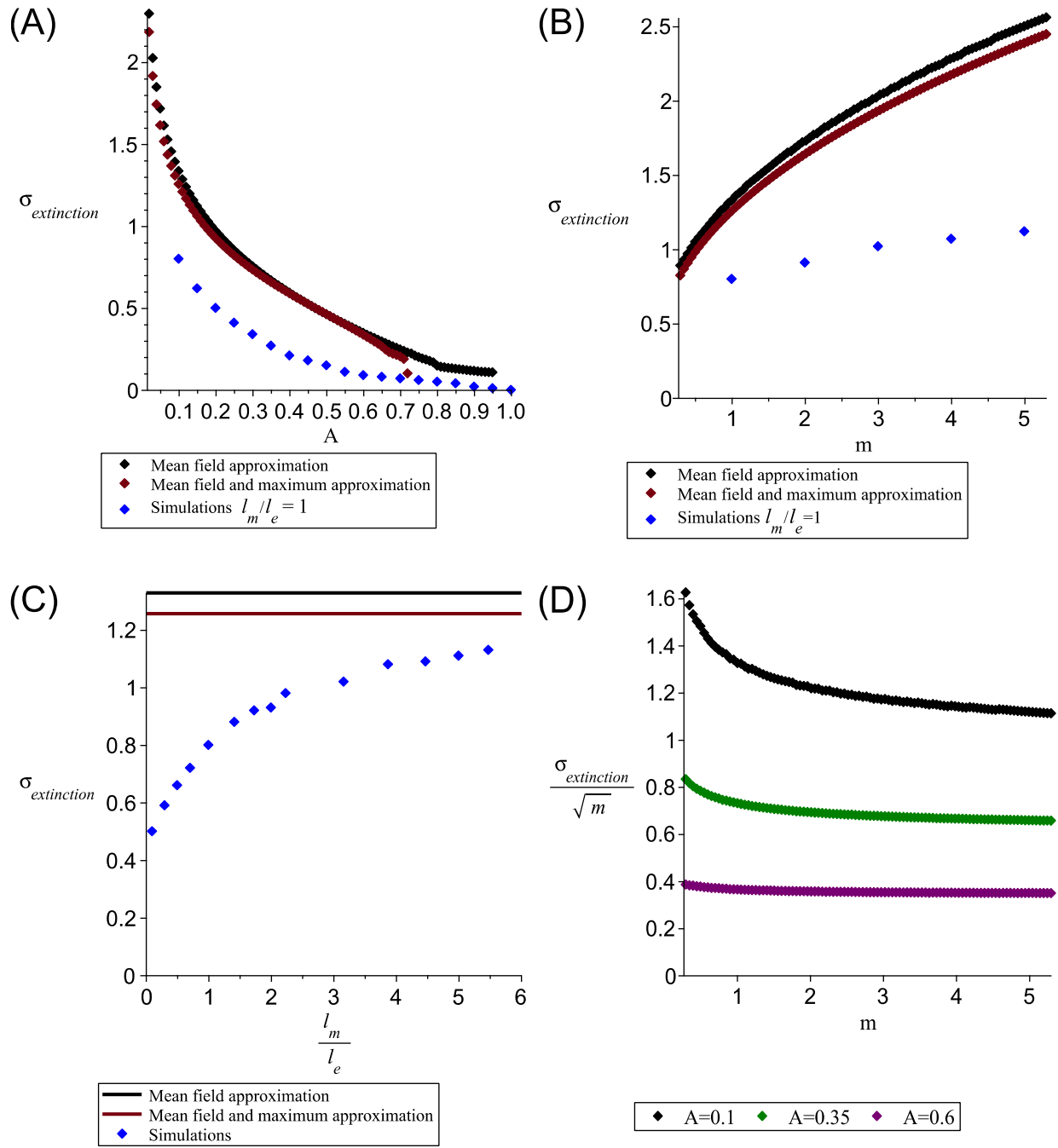

***Figure 3: Extinction threshold, $\sigma_{extinction}$, versus different parameters.*** *(Panel A) Extinction threshold versus Allee threshold for the mean-field approximation (black dots) (i.e., large dispersal length, $l_m \gg l_e$), for the mean-field and maximum approximations (red dots), and for a simulation with $l_m = l_e$ (i.e., dispersal length $l_m$ equal to the spatial scale of synchrony of environmental fluctuations $l_e$) all with dispersal rate $m=1$. (Panel B) Extinction threshold versus dispersal rate using the same color code as the previous panel, all with Allee threshold $A=0.1$.*

*The dispersal rates considered, m=0.3 to 5.3, are similar or higher than the characteristic extinction rate for these simulations, r=0.1, providing a significant contribution to population recoveries. (Panel C) Extinction threshold, $\sigma_{extinction}$, as a function of the ratio between the characteristic dispersal distance and the spatial scale of synchrony of environmental fluctuations, $l_m/l_e$, using the same color codes as the previous panels, all with m=1 and A=0.1. (Panel D) Scaling of the extinction threshold at large dispersal rate verified plotting the ratio $\sigma_{extinction}/\sqrt{m}$ versus the dispersal rate m (in the same m interval as in Panel B) for Allee thresholds A=0.1 (black), A=0.35(green), and A=0.6(purple) obtained for the mean-field approximation. For all the panels, the extinction rate is r=0.1 and the carrying capacity K=1.*

## VII. Dispersal makes the population resilient to environmental fluctuations

In a closed local population, environmental fluctuations eventually lead the population density below the Allee threshold *A*, and to extinction. Our results show that dispersal allows the recovery of a region with a depleted population thanks to population arriving from nearby non-depleted regions. This dispersal-induced population recovery makes the species resilient to population depletion caused by environmental fluctuations. Resilience to environmental fluctuations is enhanced increasing dispersal (either with larger dispersal rate or with larger dispersal length), stressing the relevance of the rescue effect of dispersal. See Fig. 3, where the extinction threshold $\sigma_{extinction}$ is represented. When environmental fluctuations are larger than the extinction threshold, the population gets globally extinct.

From the mean-field approximation, we get that in the absence of dispersal, *m=0*, the population probability, Eq. 5, diverges in *N=0* as $N^{-2\left(1+\frac{r}{\sigma^2}\right)}$, indicating that the population always goes extinct (after a certain transition time). However, when dispersal is present, the dispersal term with *I* suppresses the divergence at zero population density, and species can be sustained. Then, dispersal (or migration) is necessary to maintain a population at long times. This result is consistent with previous results found with a constant migration term in [26]. They considered a constant external migration in the growth equation, instead of the spatial extended diffusion within the habitat that we considered. In the case with dispersal, the system of equations formed by Eqs. (6) and (7) is numerically found to have either two roots or no root (different from the zero root). (See Fig. 2A.) The two regimes are separated by a critical value of the amplitude of environmental fluctuations, $\sigma_{extinction}$. (The results obtained with the mean-field approximation are shown with black dots in Fig. 3.) For values above this extinction threshold, $\sigma_{extinction}$, populations get extinct in all the locations. In contrast, for values below the extinction threshold, there is an equilibrium between extinction and recovery from extinction due to dispersal from other regions. The two branches of solutions represent two different equilibria. The high mean population density solution, $p_{high}$ (green curves in Panels B, D, and F of Fig. 1) has most regions of space with population densities above the Allee threshold, and local extinction is rare. The regions above the Allee threshold have population densities below the carrying capacity due to the cost of recovering areas with local extinction. The low mean value solution $p_{low}$ (red curves in Panels D and F of Fig. 1) has most of the regions below the Allee threshold, and they are just prevented from extinction due to the dispersal contributions from the regions with population densities above the Allee threshold. These two ideal solutions have been obtained in the mean-field approximation, which assumes large dispersal lengths *($l_m \gg l_e$)*. Nonetheless, this approximation is a limit case that can help understand populations with shorter dispersal lengths, because their extinction thresholds have similar parameter dependences (Figures 3A and 3B).

For finite dispersal length (Fig. 1), a much more common situation, simulations may present low and high-density solutions at different regions of the same habitat, together with regions of extinction (regions of zero population density in Panel E of Fig. 1, reflected as peaks at zero in the histogram of Panel F of Fig. 1).

The additional maximum approximation explained in the methods section works well for low values of A, reproducing the results of the mean-field approximation ($l_m \gg l_e$), as shown in Fig. 3.

On the one hand, the extinction threshold is found to decrease as the Allee threshold increases (Panel A of Fig. 3), as both simulations and mean-field approximation show. On the other hand, increasing the dispersal rate increases the extinction threshold, which at large values of the dispersal rate grows as

$$\sigma_{extinction} \sim \sqrt{m}, \qquad (9)$$

as we expected from Eq. (8), obtained by the mean-field and maximum approximation (See Panels B and D of Fig. 3.). This dependence of the extinction threshold with the square root of the dispersal rate is related to the stochastic nature of the environmental fluctuations (which is here modelled with a Wiener process). Finally, maximum values of the extinction threshold are found in the mean-field limit, where characteristic dispersal distance is much larger than the spatial scale of environmental synchrony, $l_m \gg l_e$ (right-hand side of Panel C of Fig. 3). For values of the dispersal distance of the order of the spatial scale of environmental synchrony, the extinction threshold is reduced (for example, to half the value given by the mean-field approximation in the simulation shown in the blue dots of Panel C of Fig. 3). Therefore, resilience to environmental noise is reduced when dispersal is less frequent (lower $m$) or more local (lower $l_m$).

## VIII.   *Discussion*

We have shown that dispersal can make a population with an Allee threshold more resilient to environmental noise-induced extinction. This resilience can be characterized by the extinction threshold $\sigma_{extinction}$ for the amplitude of environmental fluctuations, above which the species gets extinct. This extinction threshold increases if the dispersal rate $m$ increases, or if the relative dispersal length $l_m/l_e$ increases. This result is consistent with the relevance of the rescue effect of dispersal, which is proven to be an effective mechanism reducing local and global extinction risk in spatially extended populations [37,38] and has been studied in Allee effect dynamics [39,40]. The rescue effect entails that dispersal or migration can help repopulate extinct patches, reduce extinction risk, and improve the long-term sustainability of a species [41].

Mean-field approximation leads us to identify two branches of sustainable population distributions. For one of the branches, high-density population state, the population distribution has most regions above the Allee threshold and some regions below it. (See blue curve in Fig. 1F.) For the other branch, low-density population state, the population distribution has most regions with a population below the Allee threshold but sustained by dispersal from the regions above the Allee threshold. (See red curve in Fig. 1F.) Dennis et al. [26] already identified the two branches, of low and high population densities, but for one location with external migration. Here, this analysis has been done for a spatially extended population with only internal dispersal, allowing recolonization of regions with local extinction by neighboring populations. We also show that the path to global extinction is a path through the emergence of local depletions and extinctions, which spatially coexist with non-depleted regions (even for homogeneous habitats).

In this path to global extinction, low-density and extinct regions cover a more significant fraction of the area as the amplitude of environmental fluctuations increases towards the extinction threshold. Different spatial domains are close to different mean-field steady state solutions, as shown in Fig. 1 (Panels A, C, and E). Finite dispersal length decouples distant regions allowing for the coexistence of different solutions in different regions. However, the migrations between regions through the borders modifies the distribution in each region from the ideal mean field distributions. A linear combination of both mean-field distributions appears in finite dispersal simulations and in mean-field simulations (with theoretical infinitely large spatial scales of population synchrony, see Appendix A), suggesting that low- and high-density metapopulations may coexist in neighboring regions. In the mean-field limit, the zero population density solution (i.e., extinction) does not coexist with low- and high-density solutions, while for finite dispersal length, extinct regions coexist (with low and high-density solutions) for amplitudes of environmental fluctuations close to global extinction.

The typical size of the population depleted zones is given by the spatial scale of population synchrony [32,42]. The spatial scale of population synchrony arises by environmental fluctuation synchrony and can be modulated by dispersion and trophic interactions [33,43–45]. The characteristics of the spatial and temporal environmental correlations can modulate the characteristics of the transition to extinction [46]. Identifying the causes of population synchrony in natural populations is a challenge for ecologists due to the complexity and amount of data, which are sometimes incomplete and inaccurate, and thus, it is difficult to identify all the factors involved [43]. Additionally, spatial scales of population synchrony give the typical size of the areas affected by local extinction, and seem to be related to transitions to extinction in population dynamics [47–49]. In the case of infinitely long dispersal distances (i.e., mean-field approximation) we would have also infinitely large spatial scales of population synchrony, and thus, infinitely large areas with local extinction (i.e., the whole habitat). This infinite typical size of areas affected by extinction is coherent with our results in the mean-field limit, where either the population becomes globally extinct or no regions present local extinction after long times.

The low-density population state may be related to the absence of recovery seen in some ecosystems after halting harvesting [50]. In those ecosystems, the species seems to be trapped in a low-density population state, where they have been lead by harvesting. They do not recover to the previous high-density population state unless harvesting is stopped during long periods, usually taking tens of generations. Our additional simulations with an initial population in the low-population distribution shown that the low-population density state alone is unstable in our model. After enough time, it evolves towards: a high-density population state, a coexistence of regions with low- and high-population states, or global extinction, depending on the parameter values. The transitory dynamics presents moving fronts between the low and the high-population density state, which could provide clues on recovery dynamics [51–53]. Low-population density state might be a separatrix between extinction and non-zero population state, because initial conditions close to the low-density state seem to be particularly sensible to stochasticity. However, low-population density state seems to be locally stable when it is close to a high-population density state. Dennis et al. [26] also showed that external migration can stabilize the low-population density state, and lead to basin hopping between low and high-population density states. Similar behavior can be present in our model due to dispersal from a neighboring high-population region. Simulations with amplitudes of environmental fluctuations close to the extinction threshold show the stochastic emergence of extinction regions inside low-population density regions, even if initially the population density was at the carrying capacity in all points in space. (See Fig. A4 in Appendix A.) Studies concerning similar growth equations and demographic (instead of environmental) fluctuations [27,54] show a catastrophic transition between a high-density state and extinction. However, they did not find contributions of a low-density state in the stationary distribution of the population. This difference in the presence (or absence) of this contribution can be due to environmental fluctuations or the different dispersal terms, which are the main differences with our models. In addition, Villa et al. [27] suggest that smoother transitions appear in smaller spatial dimensions, and catastrophic shifts are prevented, which could explain a contribution of a low-population density state in one-dimensional models, as the one studied here.

Further studies of transitions between low and high-density population states have to be done to deeply understand the implications of the two states presented here for these ecosystems. Our current results already suggest that the repopulation of an area can lead to a change from the low to the high-density population state, if it is intense enough to lead to a change in the regime of the dynamics in this area. Further studies could provide additional clues to optimize repopulation strategies.

It would also be interesting to address the particular case where the Allee effect arises due to a population density-dependent mating rate, as in the studies with lattice models done in Refs. [55,56]. Our results have similarities (and differences) with the results obtained in these studies with lattice Monte Carlo simulations in an agent-based approach. They studied the mortality-resilience and found three equilibrium values for the mean site population in the mean-field limit: high, low and extinct. Their low population state was characterized as an unstable separatrix between the high and the extinct state in the mean-field limit. Here we have gone a step beyond and show that this low population state can be locally stable in the presence of a nearby high population state (as was in Ref. [26] by external migration). These results stimulate further research beyond the mean-field limit to get a deeper understanding of the spatially extended population dynamics in the presence of environmental fluctuations.

Additionally, the individual-based model in Ref. [57] supports the crucial role of dispersal to sustain a population. This model shows local competition or cooperation among neighbor individuals lowers the effective Allee threshold compared to global competition or cooperation. Analogously our study considers local density regulation showing that coupling through dispersal can lead to recovery of patches close or below the Allee threshold. (Note we only consider mean-field dynamics in dispersal and only for comparison with limited-range dispersal.) All these results stress the relevance of population spatial structure.

Our results compute the extinction threshold, providing an approach for assessing extinction risk under an increase of the amplitude of environmental fluctuations. (An increase in climate variability was reported for several Earth regions [29].) The numerical analysis we performed here indicates towards the low-density state being an unstable state appearing only at high amplitudes of the environmental fluctuations, close to the transition to extinction. This low-density state is locally and transitorily stabilized, probably by dispersal, and it might play a relevant role on (or be a good indicative of) the dynamics of populations close to the Allee threshold. These results seems in accordance with the three types of patches observed in a locally endangered butterfly [25], where probably this effect is enhanced and fixed in space due to habitat heterogeneities. The solutions identified, high and low-density population states, require a future more detailed analysis of stability and transitions between them and to extinction, and factors that influence these transitions, like harvesting and habitat suitability (which in nature are generally heterogeneous and stochastic in space and time).

This theoretical framework also allows studying the impact of fragmentation (i.e., the effects of finite-sized habitats), which introduces an effective limit in dispersal. Therefore, fragmentation decreases the potential of dispersion to recover populations and reduces the resilience to environmental fluctuations [58]. This analysis in the theoretical framework proposed here will provide information on the resilience of the high and low-density population states in finite-size habitats, assessing the impact of these states and transitions for the species sustainability on a scenario of increasing climate variability [29].

The model presented here could shed new light on source-sink dynamics by including external migration and variations in habitat quality (e.g., adding position dependence of the parameters) in this model [59,60]. On source-sink dynamics, some patches (sources) are more suitable and allow populations to increase, while others (sinks) have low quality and cannot sustain populations independently. However, populations in sinks can be sustained by excess individuals coming from sources. The source-sink effect has been observed in endangered populations, which present Allee Effect (in some patches) mitigated by inmigration (from other patches) [25]. Heterogeneous patches coupled through dispersal can give a non-monotonic dependence of the extinction probability with the dispersal rate [61].

*Acknowledgments*
We acknowledge Daniel Oro and Bernt-Erik Saether for their comments on the manuscript.

*Author Contributions*
RCM and FJCG conceived the manuscript. JJ wrote the original code used for the stochastic population dynamics. RCM adapted the code to populations with Allee effect. RCM ran the numerical simulations. RCM and FJCG wrote the first version of the manuscript. RCM prepared the figures. All authors discussed the results and contributed to the final manuscript.

*Funding*
This work was financially supported by 817578 TRIATLAS project of the Horizon 2020 Programme (EU). FJCG acknowledges financial support through grants: UCM-EEA-ABEL-02-2009 and 005-ABEL-CM2014A of the European Economic Area (EEA) under the NILS –

Appendix A: Supplementary figures

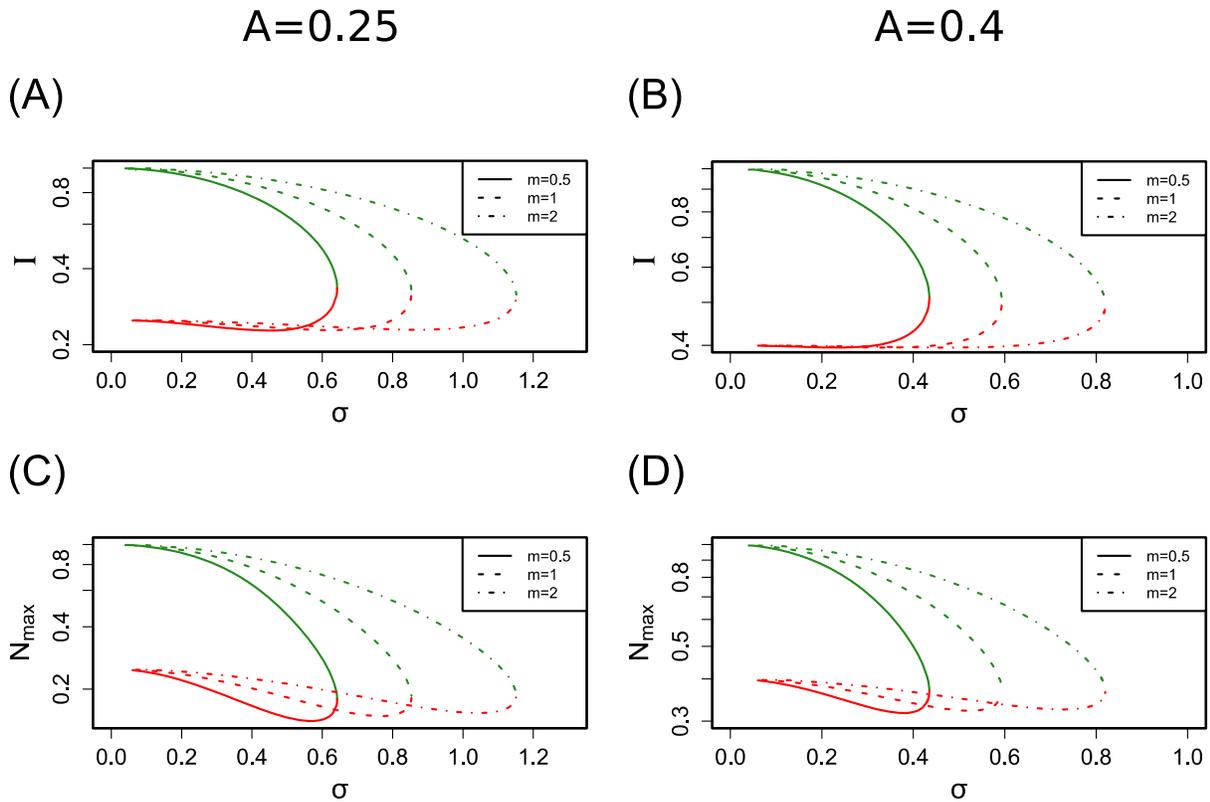

***Figure A1: Solutions in the mean-field approximation***: *Mean population density I as a function of environmental noise amplitude σ for the two nonzero branches of solutions at the mean-field approximation, with extinction rate r=0.1, carrying capacity K=1, dispersal rate m=0.5 (solid line), m=1 (dashed line) and m=2 (dash-dotted line), for Allee threshold A=0.25 (Panel A) and A=0.4 (Panel B). Position of the maximum of mean-field distributions $N_{max}$ as a function of environmental noise amplitude σ for the two nonzero branches of solutions at the mean-field approximation, with the same parameters as Panels A and B, and for Allee threshold A=0.25 (Panel C) and A=0.4 (Panel D).*

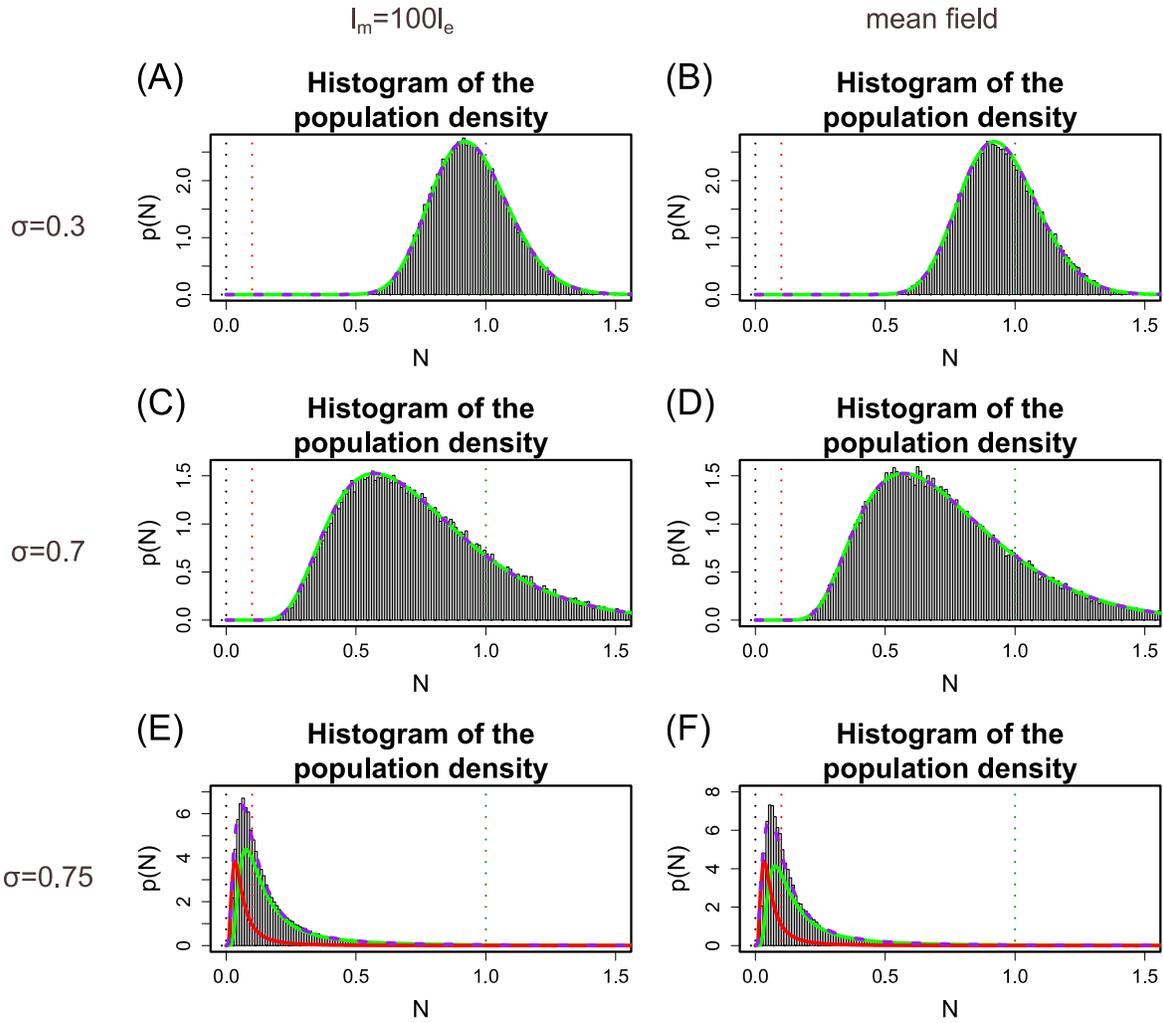

*Figure A2: Simulated population density histograms compared with mean-field population probability distributions.* Panels A, C, and E show the histograms obtained for the spatial profile of the population density for simulation at a long time, t=1000, with dispersal distance equal to 100 times the spatial scale of synchrony of environmental fluctuations, $l_m=100\, l_e$, and periodic boundary conditions. Panels B, D, and F show the histograms obtained for the spatial profile of the population density for simulation at a long time, t=1000, and mean field approximation. For every panel we considered extinction rate r=0.1, Allee threshold A=0.1, carrying capacity K=1, dispersal rate m=1, spatial scale of synchrony of environmental fluctuations $l_e=1$, and a total length of the simulation box L=4000. Panels also show the fit (purple dashed line) to a linear combination of the mean-field population probability distributions. The result for this fits are $p(N)=p_{high}(N)$ in Panels A to D, $p(N)=0.20 p_{low}(N)+0.80 p_{high}(N)$ in Panel E and p(N)=

$0.23 p_{low}(N)+0.77 p_{high}(N)$ in Panel F. Each contribution is represented with its fitted weight. *$p_{low}(N)$ (red line) and $p_{high}(N)$ (green line) correspond, respectively, to the low and high-density mean-field population probability distribution solutions. They are given by the two nonzero branches of solutions of the mean-field equations for values of σ below the extinction threshold ($σ_{extinction}$=1.33, close to σ=1.3 in Panels E and F; see also Fig. 2). Red dashed vertical lines indicate Allee threshold value A=0.1, green dashed lines indicate carrying capacity K=1 and black dashed lines indicate N=0. Note the similarities between the population density histograms obtained from direct numerical simulation and the fit to the linear combination of the population probability distributions obtained with the mean-field limit approximation.*

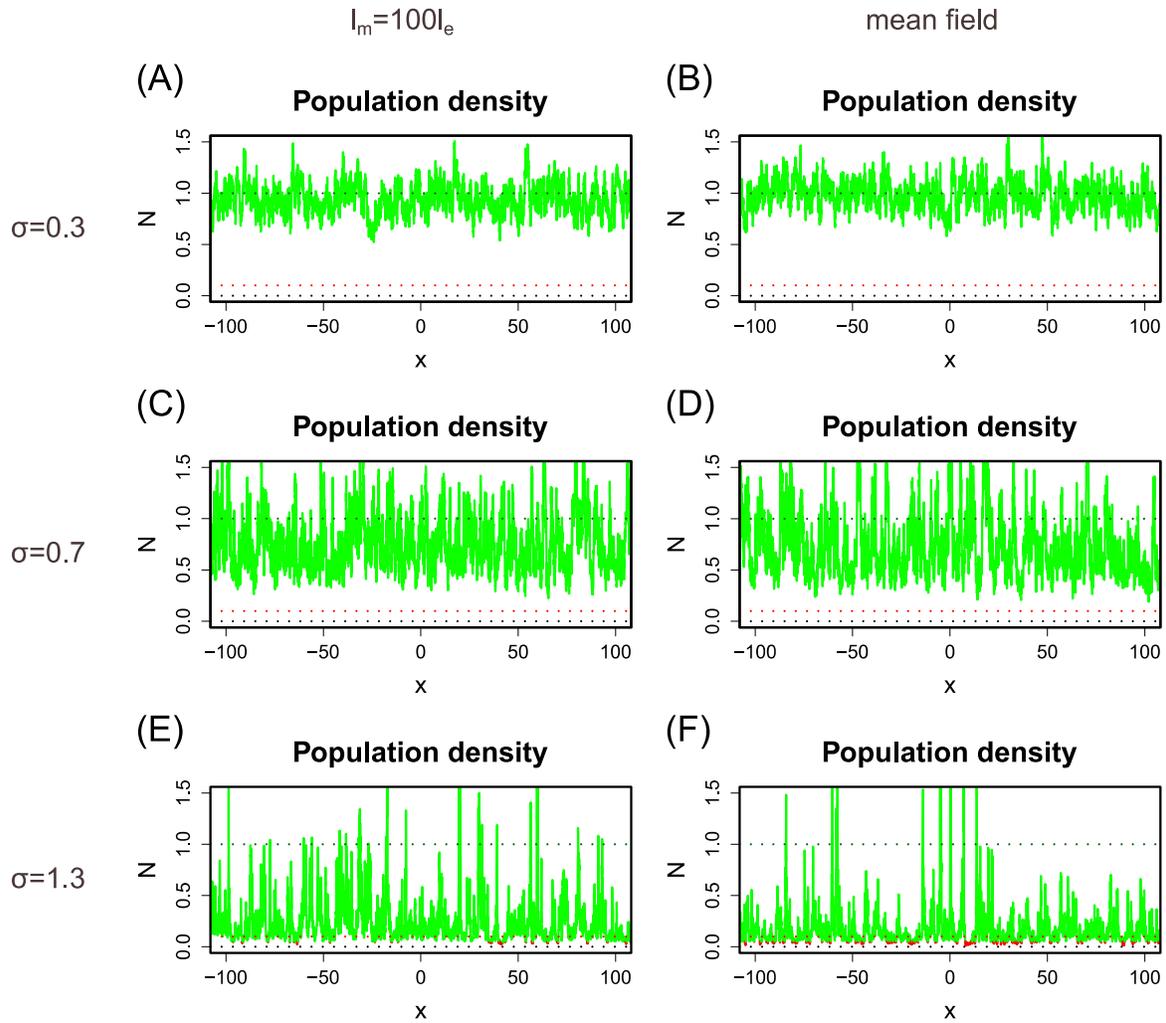

***Figure A3: Simulated population densities compared with mean-field population probability distributions, for different amplitudes of environmental fluctuations σ.*** *Panels A, C, and E*

*show the spatial profile of the population density at late time, $t = 1000 = 100r^{-1}$, with dispersal distance $l_m$=100 $l_e$, and spatial scale of synchrony of environmental fluctuations $l_e$ =1, and periodic boundary conditions. Panels B, D, and F show the spatial profile of the population density for simulation at a late time, $t = 1000 = 100r^{-1}$, and mean field approximation. Every panel gives the associated histogram in the same panel in Figure A2. For every panel we considered extinction rate r=0.1, Allee threshold A=0.1, carrying capacity K=1, dispersal rate m=1, spatial scale of synchrony of environmental fluctuations $l_e$=1, and a total length of the simulation box L=4000 (we only represent from x=-100 to x=100 to improve visualization of spatial structure). The curves represented show the patches high-population (green), and low-population (red) states according to which distribution dominates in the respective panel of figure A2. Here, σ$_{extinction}$=1.33, close to σ=1.3 in Panels E and F; see also Fig. 2. Red dashed vertical lines indicate Allee threshold value A=0.1, green dashed lines indicate carrying capacity K=1 and black dashed lines indicate N=0.*

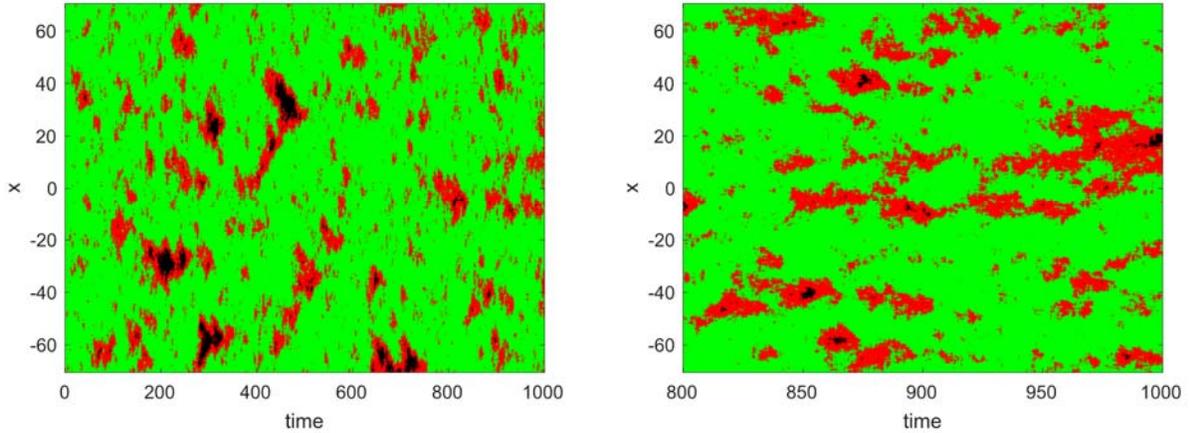

*Figure A4: Evolution of the spatial profile of population density from $t = 0$ to $t = 1000 = 100r^{-1}$ (left), and zoom on the late evolution from $t = 800 = 80r^{-1}$ to $t = 1000 = 100r^{-1}$ (right). Extinction rate $r = 0.1$, Allee threshold $A = 0.1$, carrying capacity $K = 1$, dispersal rate $m = 1$, dispersal distance equal to the spatial scale of synchrony of environmental fluctuations, $l_m = l_e = 1$, and amplitude of the environmental fluctuations $\sigma = 0.75$ (while the extinction threshold is $\sigma_{extinction} = 0.80$). The colors in the figure correspond to patches in high-population (green, N>0.1), low-population (red, 0.01<N≤0.1), and extinction states (black, N≤0.01), the criterion used in Fig. 1E. These states correspond to high-density, low-density, and extinction distribution in Fig. 1F.*

## Appendix B: Time to extinction in the absence of dispersal

The stationary population density distribution, $p(N)$, without dispersal is given by Eq. 5. This equation shows that the population distribution in the absence of dispersal is not normalizable, because it diverges at zero population density with $N^{-2(1+r/\sigma^2)}$. Therefore, the population will die out at long times. However, Eq. 5 does not indicate when this extinction will occur. Divergence at zero being greater for smaller environmental fluctuations may suggest that extinction happen faster for lesser environmental variability, but this is not the case as we show below.

Mean first passing time [26,62] can be used to study the expected time to extinction of the population. Hence, the expected time taken for a population starting at a specific initial population density $N_0$ to reach a smaller population density $N_f$ is described by

$$\tau_{N_0 \to N_f} = 2 \int_{N_f}^{N_0} \frac{\int_N^\infty p(z)dz}{v(N)p(N)} dN, \quad (B1)$$

where $p(N)$ is described by Eq.5, $v(N) = \sigma^2 N^2$, and $N_0$ is the initial population density. Replacing $p(N)$ and $v(N)$ we obtain

$$\tau_{N_0 \to N_f} = \frac{2}{\sigma^2} \int_{N_f}^{N_0} N^{2r/\sigma^2} \frac{\int_N^\infty \left( \frac{\exp\left(\frac{1}{\sigma^2}\left(\frac{2rz}{K} + \frac{2rz}{A} - \frac{rz^2}{AK}\right)\right)}{z^{2+2r/\sigma^2}} \right) dz}{\exp\left(\frac{1}{\sigma^2}\left(\frac{2rN}{K} + \frac{2rN}{A} - \frac{rN^2}{AK}\right)\right)} dN. \quad (B2)$$

Exact time to extinction is reached when the chosen extinction threshold reference $N_f$ tends to zero. However, the numerical computation becomes very slow in this case, and we chose instead a small nonzero value of $N_f$. For example, the Allee threshold $A$, or populations 10 or 100 times smaller, which in many cases is equivalent to extinction (because it is of the order of one or few individuals, unable to survive). Fig. B1 shows that the time needed to reach such a small population (starting from the carrying capacity K) decreases with the amplitude of environmental fluctuations σ. Smaller final population sizes $N_f$ imply slightly longer first passing times. However, considering for $N_f$ the Allee threshold $A$, or 100 times less individuals yield almost no difference in the extinction time.

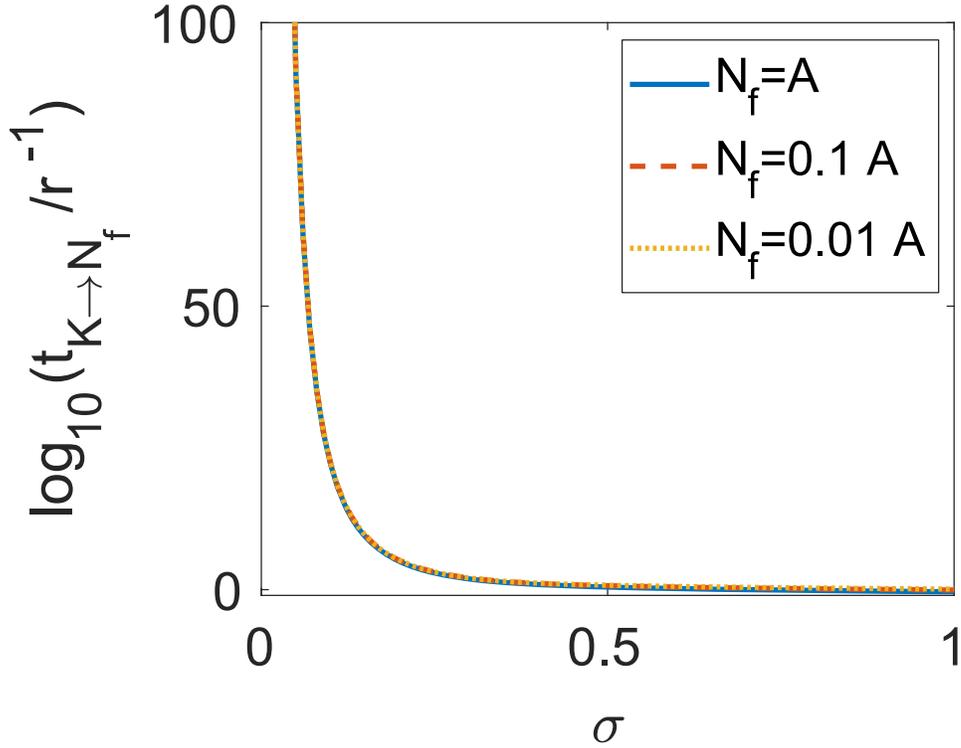

*Figure B1: Approximation of the extinction time in the absence of dispersal. Decimal logarithm of the expected time taken for a population starting from the carrying capacity K to reach a final population $N_f$, divided by the inverse extinction rate $r^{-1}$. We have considered a carrying capacity K=1, Allee Threshold A=0.1, and a extinction rate r=0.1. We can see in the figure that choosing the Allee Threshold A as final population density yield a very similar extinction time than choosing a final population density 100 times smaller.*

## Appendix C: Simulation algorithm

We begin by setting the parameters for the simulation: Allee threshold $A$, carrying capacity $K$, extinction rate $r$, migration rate $m$, mean dispersal distance $l_m$, the amplitude of environmental fluctuations $\sigma$, and spatial scale of environmental synchrony $l_e$.

Space and time are discretized. The spatial grid is an array of length $n$ (a natural, odd number) from $x = -L_{end}$ to $x = L_{end}$ (representing a box of size $2L_{end}$). The characteristic spatial scales are the dispersal distance $l_m$, and the spatial scale of synchrony of the environmental fluctuations $l_e$. We chose $L_{end}$ (at least) 20 times the larger characteristic time scale. (For example, 20 times $l_m$ in Figs. A2 and A3, where $l_m = 100 l_e$, while for most of the other simulations 70 times the maximum spatial scale.) The spatial vector length $n$ is chosen such as the distance between neighbor nodes $\Delta L = 2L_{end}/(n-1)$ is (at least) 20 times smaller than the minimum spatial scale. This spatial resolution and box size has been shown to provide an accurate description of the population dynamics in an infinite habitat (results are independent of $L_{end}$ and $\Delta L$).

The temporal grid is an array from $t = 0$ to $t = t_{end}$, with a distance between nodes of $\Delta t$, which is taken as 50 times the minimum temporal scale, and $t_{end}$ is 100 times the maximum temporal scale. The characteristic temporal scales are $1/r$ and $1/m$.

The initial conditions at $t = 0$ are all spatial nodes with a population density equal to the carrying capacity, $N(x_i, t = 0) = K$, unless stated otherwise.

Once the grid has been defined, and the initial conditions set, we begin the simulation. The differential equation that governs the model is defined by Eq. (3). To implement it numerically, we can calculate the population density at a specific point in time and space, by using the Euler Algorithm such as

$$N(x_i, t_{j+1}) = N(x_i, t_j) + \Delta N(x_i, t_j)\Delta t + \sigma N(x_i, t_j)\zeta(t_j)\sqrt{\Delta t} \qquad (C1)$$

Where $\zeta(t_j)$ is an exponentially autocorrelated Gaussian Field with zero mean, variance equal to 1, and correlation distance equal to $l_e$, and $\Delta N(x_i, t_j)$, the deterministic contribution, is

$$\Delta N(x_i, t_j) = r \cdot N(x_i, t_j) \left(\frac{N(x_i, t_j)}{A} - 1\right)\left(1 - \frac{N(x_i, t_j)}{K}\right) \cdot \Delta t - m \cdot N(x_i, t_j) \cdot \Delta t$$

$$+ m \cdot \Delta t \cdot \sum_{k=-\frac{n-1}{2}}^{\frac{n-1}{2}} N(x_{i-k}, t_j) \cdot \frac{1}{\sqrt{2\pi l_m^2}} e^{\frac{-(k \cdot \Delta L)^2}{2 l_m^2}} \cdot \Delta L \qquad (C2)$$

Note that in the sum $x_{i-k}$ can have values outside of the spatial grid. We can solve that by setting periodic boundary conditions, which means that the first and the last point of the grid behave as neighboring nodes, implying that an individual which disperses beyond the last patch appears at the beginning of the grid (and vice-versa), i.e.:

$$N(x_{i-k}, t_j) = \begin{cases} N(x_{i-k+n}, t_j) \text{ if } 1 > i - k \\ N(x_{i-k}, t_j) \text{ if } 1 \leq i - k \leq n \\ N(x_{i-k-n}, t_j) \text{ if } i - k > n \end{cases} \qquad (C3)$$

In case we want to simulate the dispersal dynamics in the mean-field limit (as done in the simulations for Figures 2, A2 and A3), we change the last term of Eq. (C2), the dispersal term, and we use instead

$$\Delta N(x_i, t_j) = r \cdot N(x_i, t_j) \left(\frac{N(x_i, t_j)}{A} - 1\right)\left(1 - \frac{N(x_i, t_j)}{K}\right) \cdot \Delta t - m \cdot N(x_i, t_j) \cdot \Delta t$$
$$+ m \cdot \Delta t \cdot \frac{\sum_{k=1}^{n} N(x_k, t_j)}{n}. \quad (C4)$$

(For mean-field simulations we have set $L_{end} = 2000 \, l_e$.)

We have verified the results for these algorithms with different spatial and temporal resolutions and we found the results are accurate, stable and consistent.